\documentclass[twocolumn,prl]{revtex4-1}
\usepackage{graphicx}
\usepackage{amssymb}
\usepackage{epstopdf}
\usepackage{amsmath,bm}
\usepackage{footnote}
\usepackage[bottom]{footmisc}
\usepackage{color}
\makeatletter

\@ifundefined{textcolor}{}
{
 \definecolor{BLACK}{gray}{0}
 \definecolor{WHITE}{gray}{1}
 \definecolor{RED}{rgb}{1,0,0}
 \definecolor{GREEN}{rgb}{0,1,0}
 \definecolor{BLUE}{rgb}{0,0,1}
 \definecolor{CYAN}{cmyk}{1,0,0,0}
 \definecolor{MAGENTA}{cmyk}{0,1,0,0}
 \definecolor{YELLOW}{cmyk}{0,0,1,0}
}

\usepackage[colorlinks,
            linkcolor=black,
            anchorcolor=black,
            citecolor=black
            ]{hyperref}
\usepackage{ulem}           
            
\makeatother

\begin{document}

\title{Topological proximity effects in a Haldane-graphene bilayer system}

\author{Peng Cheng$^{1,\dagger}$, Philipp W. Klein$^{1,\dagger}$, Kirill Plekhanov$^{1,2,3}$,  Klaus Sengstock$^{4,5,6}$, Monika Aidelsburger$^{7,8}$, Christof Weitenberg$^{4,5}$,  Karyn Le Hur$^{1,*}$}
\affiliation{$^1$  Centre de Physique Th\' eorique, \'Ecole Polytechnique, CNRS, Universit\'e Paris-Saclay, Route de Saclay, 91128 Palaiseau, France} 
\affiliation{$^2$ LPTMS, CNRS, Univ. Paris-Sud, Universit\' e Paris-Saclay, 91405 Orsay, France}
\affiliation{$^3$ Department of Physics, University of Basel, Klingelbergstrasse 82, CH-4056 Basel, Switzerland}
\affiliation{$^4$ ILP --- Institut f\"ur Laserphysik, Universit\" at Hamburg, Luruper Chaussee 149, 22761 Hamburg, Germany}
\affiliation{$^5$ The Hamburg Centre for Ultrafast Imaging, Luruper Chaussee 149, 22761 Hamburg, Germany}
\affiliation{$^6$ Zentrum f\" ur Optische Quantentechnologien, Universit\" at Hamburg, 22761 Hamburg, Germany}
\affiliation{$^7$ Fakult\" at f\" ur Physik, Ludwig-Maximilians-Universit\" at, Schellingstrasse 4, 80799 M\" unchen, Germany}
\affiliation{$^8$ Max-Planck-Institut f\"ur Quantenoptik, Hans-Kopfermann-Str. 1, 85748 Garching, Germany}
\date{\today}

\begin{abstract}

We reveal a proximity effect between a topological band (Chern) insulator described by a Haldane model and spin-polarized Dirac particles of a graphene layer.
Coupling weakly the two systems through a tunneling term in the bulk,  the topological Chern insulator induces a gap and an opposite Chern number on the Dirac particles at half-filling resulting in a sign flip of the Berry curvature at one Dirac point.  We study different aspects of the bulk-edge correspondence and present protocols to observe the evolution of the Berry curvature as well as two counter-propagating (protected) edge modes with different velocities. In the strong-coupling limit,  the energy spectrum shows flat bands. Therefore we build a perturbation theory and  address further the bulk-edge correspondence. We also show the occurrence of a topological insulating phase with Chern number one when only the lowest band is filled. We generalize the effect to Haldane bilayer systems with asymmetric Semenoff masses. We propose an alternative definition of the topological invariant on the Bloch sphere. 

\end{abstract}
\maketitle

Topological systems have attracted a considerable attention these last decades \cite{Andrei,Reviews} as they show robust gapless edge modes which are relevant for quantum information purposes \cite{RMP}. The Haldane model \cite{Haldane} on the honeycomb lattice, which has been realized in ultra-cold atoms \cite{ETH,Hamburg}, graphene \cite{Cavalleri}, quantum materials \cite{Stanford}, and photonic topological systems  \cite{HaldaneRaghu,MITlight,Circ,RevueQED,Revue} now appears as a paradigmatic model in the topological classification of Bloch energy bands. For spinless fermions, the bulk state is insulating at half-filling and characterized by a topological invariant, the first Chern number, while the edges of the system reveal a one-dimensional gapless chiral mode by analogy with the quantum Hall effect \cite{Pepper,Kim,Halperin,Buttiker,TKNN,Kohmoto}. Topological proximity effects induced by a topological band insulator \cite{TIhybrid,LeonTaylor,Frankfurt} have also started to gain interest as a generalization of the proximity effect induced from a superconductor onto a metallic system \cite{KLH,KSC}. In this Letter, we study the proximity effect when tunnel coupling a Haldane model with a layer of graphene \cite{RMPgraphene,Hamburggraphene,ETHZurich}. We assume spinless particles in both layers and
the tunnel process couples the same sublattice in the two layers.  Particle-hole processes at the interface open a gap as a result of pseudo-spin effects, inducing an inverse topological order in the graphene system when both layers are half-filled. 

The Haldane model and graphene layers are described through the same pseudospin-1/2 representation in momentum space, as a result of the two sublattices of the honeycomb lattice \cite{Cayssol},  allowing us to describe the proximity effect in the same torus representation of the first Brillouin zone and fiber bundle approach on the Bloch sphere. We address different geometries and protocols to describe the bulk-edge correspondence and the Berry curvatures \cite{Berry} of Bloch bands which could be equivalently probed for fermions and bosons at the one-particle level. We draw an analogy with the Kane-Mele model \cite{KM,Zhang,Murakami} and with the quantum spin Hall effect \cite{QSH,MIT}, regarding the edge structure. We also suggest implementations in graphene bilayers, cold atoms and light systems. 

We start our analysis with the Hamiltonian ${\cal H}={\cal H}_g + {\cal H}_h + {\cal H}_r$, where ${\cal H}_g$ describes the graphene layer, ${\cal H}_h$ the topological Haldane model, and ${\cal H}_r$ the tunnel coupling between the layers with amplitude $r$.  We emphasize here that we consider no displacement in the stack of the two layers. We use the definitions where $t_1$ means nearest-neighbor hopping element on the honeycomb lattice, $t_2$ second nearest neighbor tunneling element with the associated phases $\pm \Phi$ for sublattices $A$ and $B$ \cite{SM}. In wave-vector space, the Hamiltonian takes the form ${\cal H}=\int d{\bf k}/(2\pi^2) {\cal H}({\bf k})$, where
\begin{eqnarray}
   {\cal H}({\bf k}) =
  \left({\begin{array}{cc}
 {\bf d}_g\cdot \bm{\sigma}  & r\cdot{\mathcal{I}} \\
   r \cdot{\mathcal{I}} & {\bf d}_h\cdot \bm{\sigma} + \epsilon_h\cdot{\mathcal{I}} \\
  \end{array} } \right)
\end{eqnarray}
with the pseudo-spin Pauli matrices {\boldmath$\sigma$} acting in the Hilbert space of sublattice A and sublattice B of each layer $g$ and $h$,  respectively \cite{Cayssol}. To make an analogy with two 1/2 spins in ${\bf k}$-space, one
could also choose to introduce two different sets of Pauli matrices $\bm{\sigma}_1$ and $\bm{\sigma}_2$; the results derived below can be simplified in notations through the introduction of one set of Pauli matrices.
The magnetic fields: $d^h_x({\bf k}) = -t_1 \sum_{i=1}^3 \cos({\bf k}\cdot \mathbf{a}_i)$, $d^h_y({\bf k}) = 
- t_1 \sum_{i=1}^3 \sin({\bf k}\cdot \mathbf{a}_i)$, $d^h_z({\bf k})= -2t_2\sin\Phi \sum_{i=1}^3 \sin({\bf k}\cdot {\bf b}_i)$.  The vectors $\mathbf{a}_i$ and $\mathbf{b}_i$ link nearest neighbors and next-nearest-neighbors on the honeycomb lattice \cite{SM}. Furthermore, 
 $\epsilon_h=-2t_2\cos\Phi \sum_{i=1}^3 \cos({\bf k}\cdot {\bf b}_i)$ and ${\cal I}$ is the $2\times 2$ identity matrix. Since we assume that the nearest-neighbor tunneling amplitudes are identical in both layers (for the simplicity of notations but without loss of generality), then $d_x^g({\bf k})=d_x^h({\bf k})$ and $d_y^g({\bf k})=d_y^h({\bf k})$, and initially for graphene (when $r=0$) the magnetic field in {\bf k}-space resides in the equatorial plane $d_z^g({\bf k})=0$. In the numerical calculations below, we fix the phase $\Phi=\pi/2$. 
  
Mapping the first Brillouin zone on a torus onto the sphere $S^2$, the Haldane model at $r=0$ is characterized by the normalized magnetic field $d^*=(\sin\theta({\bf k})\cos\phi({\bf k}),\sin\theta({\bf k})\sin\phi({\bf k}),\cos\theta({\bf k}))$ such that the Chern number associated with the two bands of the topological Haldane insulator can be defined as 
\begin{equation}
C_{\pm}^h = \frac{1}{2\pi} \int_{S^2} F_{\pm} = \mp \frac{1}{4\pi}\int_{S^2} d\Omega = \mp 1,
\end{equation}
with the relation between the Berry curvature and the solid angle on the sphere $S^2$:   $F_{\pm}=\mp \sin\theta d\theta d\phi = \mp \frac{d\Omega}{2}$. In Fig. 1 top left, we show the
Berry curvature associated with the lowest energy band of the Haldane model, corresponding to the Chern number $C_-^h=C_1=+1$. The Chern number of such spin-1/2 models on the sphere $S^2$ has been measured in circuit Quantum electrodynamics by applying a one-dimensional path on the Bloch sphere going from north to south poles \cite{SantaBarbara,Boulder,LoicPeter}. The Berry curvature of the Haldane model has also been reconstructed in cold atoms \cite{Hamburg} through momentum space density which is obtained from time of flight images, $n({\bf k})=f({\bf k})[1-\sin\theta({\bf k})\cos\phi({\bf k})]$, where $f({\bf k})$ corresponds to the broad envelope associated with the momentum distribution of the Wannier function \cite{Eckardt}.  To measure accurately the two angles, one can create a chemical potential offset between the two sub-lattices $\Delta_{AB}$, which then acts in the quasi-momentum representation as a rotation (in ${\bf k}$ space), $\phi({\bf k})\rightarrow \phi({\bf k}) + \Delta_{AB} t/\hbar$ where $\hbar=h/(2\pi)$ is the (reduced) Planck constant \cite{Eckardt,Hamburg}.  Topology of the Bloch bands can also be accessed through Wilson line measurements \cite{Munich} and coupling with circularly polarized light \cite{Cavalleri,Nathan}. 

\begin{figure}[t]
   \includegraphics[width=5.5cm]{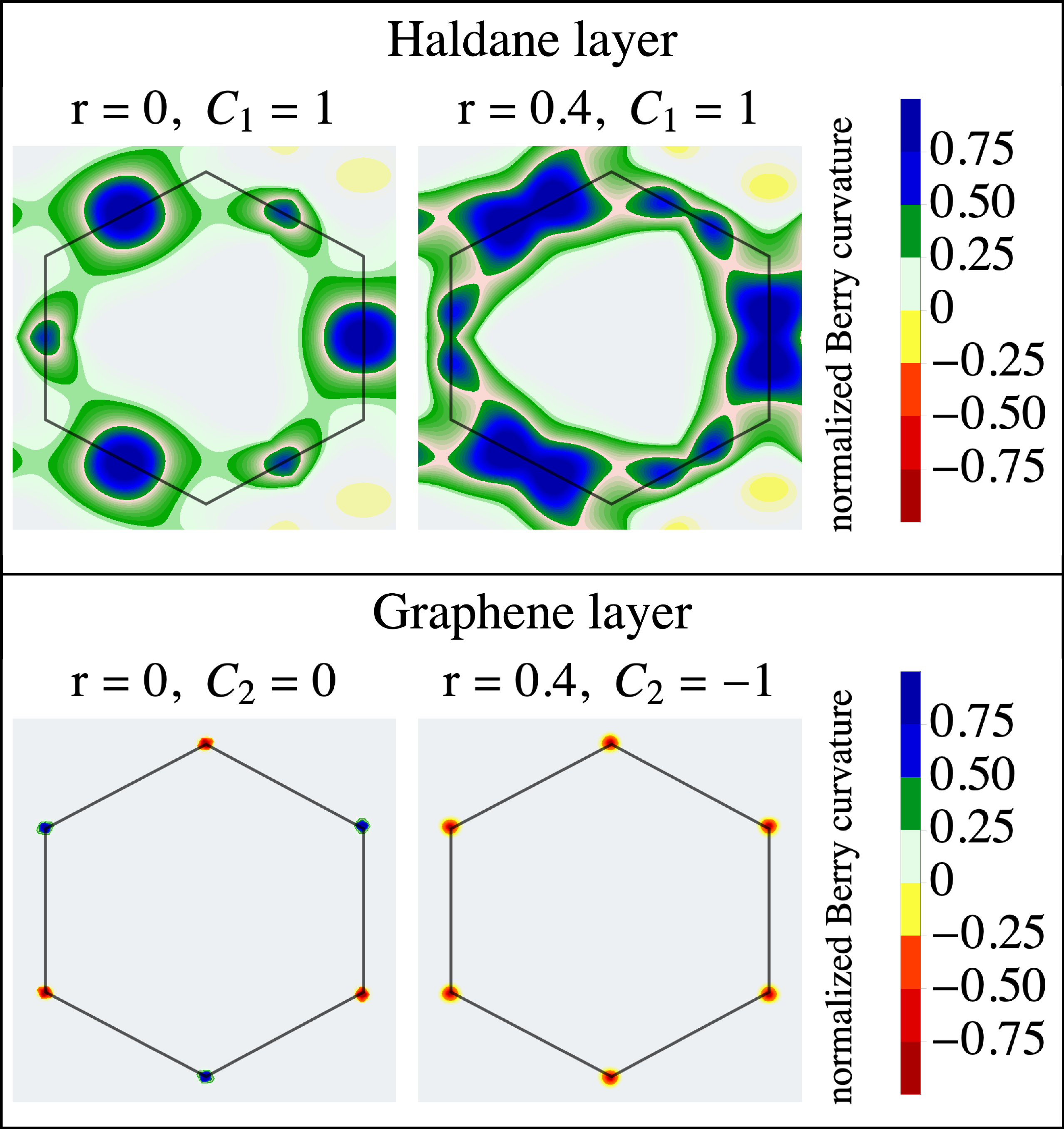}
      \caption{Berry curvature in the Brillouin zone for the Haldane and graphene layers at $r=0$ and small $r$, showing the Berry phase jump effect \cite{SM}. Here, $t_1=1$ and $t_2=1/3$.}
          \label{}
    \vskip -0.5cm
\end{figure}

The Chern number of the graphene system is equal to $C_{\pm}^g=C_2=0$ in the absence of coupling with the topological layer, {\it i.e.}, $r=0$.  One can still define a Berry phase \cite{Berry} $\pm \pi$ associated with local pseudo-spin effects in ${\bf k}$-space when linearizing the band structure around the two inequivalent Dirac points (see Fig. 1) \cite{RMPgraphene}. To show how an effective $d_z^g$ magnetic field component can be induced in the graphene layer through the presence of the $d_z^h$ term in the Haldane system, we build a path integral approach in the small $r\ll (t_1, t_2)$ limit  integrating out degrees of freedom of the Haldane model.  Assuming that the $r$ tunneling term couples mostly the same sublattice of the different layers (leading to ${\cal H}({\bf k})$ in Eq. (1)) then the partition function of the graphene layer becomes:
\begin{eqnarray}
\hskip -0.2cm Z_g &=& \int {\cal D}\zeta_g({\bf k}) {\cal D}\bar{\zeta}_g({\bf k}) \exp-\big(\int_0^{\beta} d\tau \int \frac{d^2 k}{2\pi^2} \bar{\zeta}_g({\bf k}) \big[\partial_{\tau} \\ \nonumber
&+& {\bf d}_g({\bf k})\boldmath{\sigma} - \frac{r^2}{|{\bf d}_h({\bf k})|^2}(1-e^{-\epsilon \tau}){\bf d}_h({\bf k})\bm{\sigma}\big](\zeta_g({\bf k}))^T\big),
\end{eqnarray} 
with $\zeta_g({\bf k})=(c_{gA}({\bf k}), c_{gB}({\bf k}))$ describing an electron annihilation operator in the graphene layer, at sublattice $A$ and $B$ respectively, and $\epsilon$ an energy scale close to $t_2$. 
At long time scales $\epsilon\tau\gg 1$ or low energy compared to the Haldane gap we find that the proximity effect here results in the induction of a finite magnetic field in the graphene layer along the $z$ direction. Such an effective term in the Hamiltonian agrees with second-order perturbation theory in the tunnel coupling $r$ \cite{LeonTaylor}, and appears independently of the statistics of particles or fields. 
It can also be seen as an analogue of an antiferromagnetic Ising term between two 1/2 spins in ${\bf k}$-space. 

We build an effective low-energy model close to the Dirac points of graphene $(d_x=d_y\sim 0)$, approximating $d_z^h({\bf k}\sim \pm {\bf K})=\pm 3\sqrt{3}t_2\sin \Phi$ where the proximity effect is more important at small $r$. The induced term $\sim -r^2/(27 t_2^2\sin^2\Phi)d^h_z({\bf k}){\sigma}_z$ opens a gap at the two Dirac points and builds an analogy with the Haldane model. 
This now ensures that the low-energy band of the graphene layer satisfies the following condition on the Chern number $C_-^g = C_2 =- 1$.  

\begin{figure}[t]
   \includegraphics[width=8.2cm]{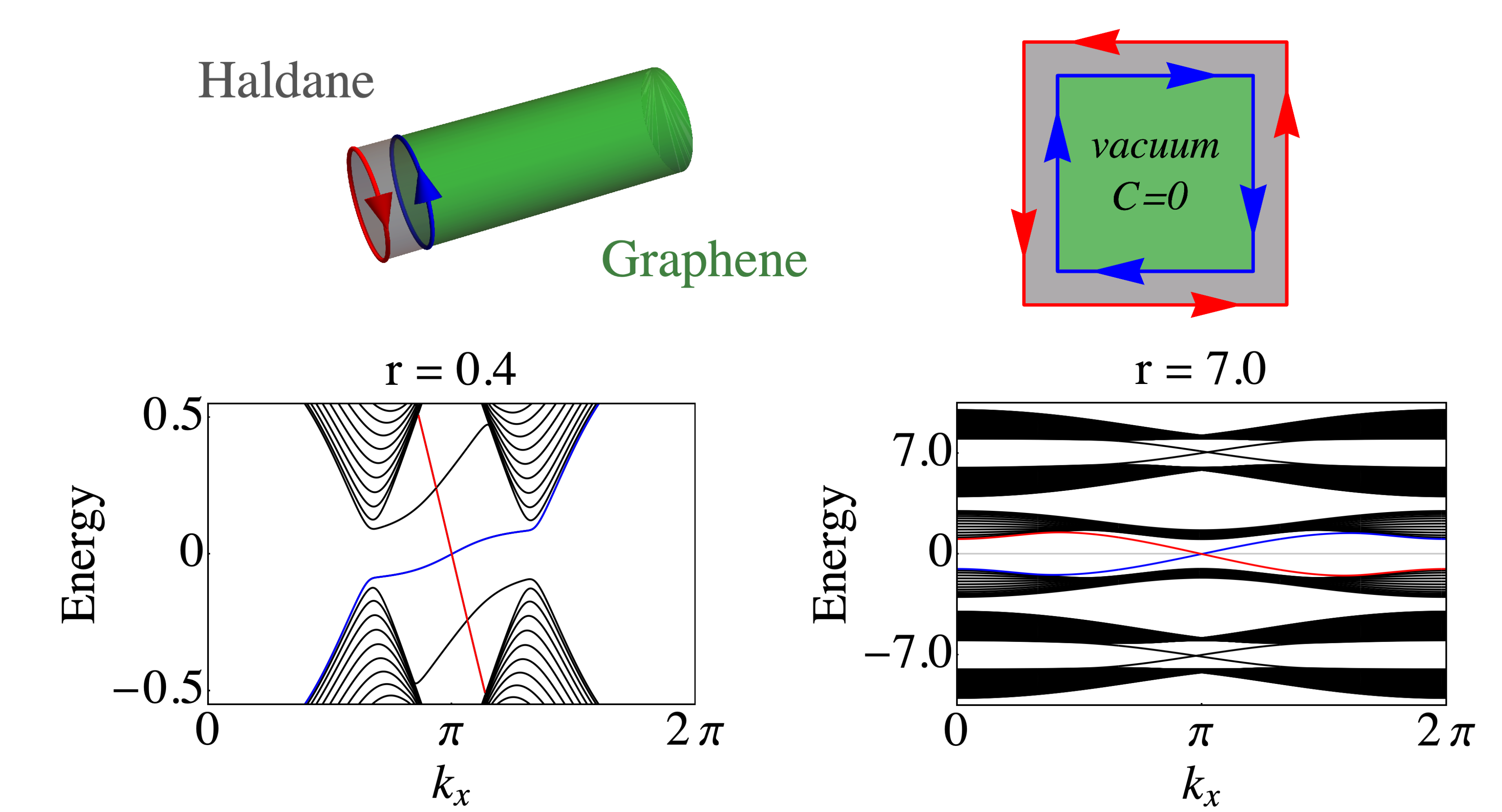}
          \label{fig:Berry}
          \caption{Band structures for $t_1=1$ and $t_2=1/3$ in the weak and strong-coupling limits for a cylinder geometry \cite{Andrei}; the lattice spacing is $a=1$. On the left, we zoom on the two low-energy graphene bulk bands. In the green region of 60 unit cells, the total Chern number of the two lowest bands is zero and in the grey region (of $14$ unit cells) the system is a Haldane model. We observe two counter-propagating edge modes with different velocities at zero energy until $r\sim t_2$. For very strong couplings, at zero energy, the counter-propagating edge modes are only linked to properties of the Haldane region.}
     \vskip -0.5cm
\end{figure}

To understand this key point, we use the form of the eigenstates close to a given Dirac point in graphene \cite{RMPgraphene}
\begin{eqnarray}
\frac{1}{\sqrt{2}}\left ( \begin{array}{c}
1 \\
\pm e^{i \phi({\bf q})} \\
\end{array} \right )
\end{eqnarray}
where ${\bf q}$ corresponds to a small deviation from a Dirac point such that $\tan\phi = q_y/q_x$.  The $\pm$ signs refer to positive energy and negative energy bands respectively meeting at the Dirac point; these two bands
are related through $\phi \leftrightarrow \phi+\pi$. It is constructive to introduce the mass (gap) $m=-r^2/(27\sin^2\Phi t_2^2) d_z^h({\bf k})$ which changes of sign at the two Dirac points in the graphene layer.  In the Supplementary Material, we provide the
forms of the eigenstates as a function of $v_F|{\bf q}|$ and $m$ \cite{SM}. We check that eigenstates converge to those of Eq. (4) in the limit $|m|\ll v_F|{\bf q}|$, where $v_F=3 t_1 a/2$ is the Fermi velocity and $a$ the lattice spacing. The two Dirac points ${\bf K}$ and ${\bf K}'$ are now related through a mass $m\rightarrow -m$ inversion corresponding to change $\phi \rightarrow \phi' =-\phi+\pi$.  The operation $\phi\rightarrow -\phi$ corresponds to change ${\bf K}$ in ${\bf K}'$ and the additional $\pi$ phase corresponds to invert the upper and lower bands. We find that the Berry phases at the two Dirac points become equal \cite{SM}
\begin{equation}
\gamma_K = \gamma_{K'} = -\frac{1}{2}\oint {\boldmath{\nabla}}\phi({\bf k}) \cdot d{\bf k} = -\pi.
\end{equation}
The integration follows a closed path around a Dirac point. 
We numerically check \cite{NumericsBerry} that a $-\pi$ Berry phase occurs at both Dirac points of graphene (see Fig. 1), similarly to the Haldane model when $t_2\ll t_1$. We also check that for the upper band of graphene, $C_+^g=+1=-C_+^h$ or $\gamma_K=\gamma_{K'}=\pi$ (this is equivalent to change $m\rightarrow -m$ at a Dirac point or $d_z^h\rightarrow -d_z^h$; see also Eq. (19) of the Supplementary Material in Ref. \cite{SM}). This effective model could be perhaps realized in a bilayer graphene by applying circularly polarized light, then opening a Haldane gap in one layer \cite{Cavalleri}. If this gap is larger than the tunnel coupling, then one could re-write the effective tunnel coupling at the Dirac points justifying this low-energy model. Below, we shall address a generalized bilayer Haldane model which can be realized in cold atoms. 

The Berry phases could be directly measured \cite{Hamburg,Munich}.  Information on Berry phases could also be reconstructed from quantum Hall conductivity \cite{TKNN} quantum circular dichroism by shining with light \cite{Nathan}, scanning probe \cite{Burkard,Westervelt} and Klein paradox \cite{Katsnelson,DavidBenjamin} measurements. 

In the Haldane layer with $t_2 \sim t_1$, the  pseudo-spin 1/2 is polarized close to the Dirac point, and the structure of the Berry curvature is strongly modified: its dominant contribution occurs close to the highly-symmetric M points now \cite{SM}. Results in the Haldane layer remain almost unchanged from $r=0$ to $r=0.4$ (see Fig. 1).

Now, we study in more detail the edge properties. For two layers of equal size, for $r\neq 0$, we find the formation of a gap at the edges at half-filling, resulting from the hybridization of the zig-zag edge mode of graphene --- present at $r=0$ --- with the topological edge mode (see black edge modes in Fig. 2 left corresponding to the right boundary of the green cylinder). This is also consistent with the Kane-Mele model \cite{FuKane}, where the $r$ coupling at the edges corresponds to a spin flip process which breaks the ${\cal Z}_2$ symmetry and opens a gap  similar to the effect of the Mott transition \cite{Wei,Stephan}. To confirm that a chiral edge state has now appeared in the graphene layer at half-filling moving in opposite direction as the edge state in the Haldane layer, in agreement with $(C_-^h - C_-^g)=2$ in the bulk for $r\neq 0$ \cite{Sheng}, we suggest to suppress smoothly the $r$ tunnel coupling at the left edge. In the numerics, we check that for more than 10 unit cells in the grey region, results are stable: Fig. 2 then shows two counter-propagating edge modes with different velocities, due to the different gaps in the two layers, crossing the chemical potential at half-filling (or energy zero). One could build a slightly smaller layer and observe two counter-propagating edge modes, one in each layer.  

\begin{figure}[t]
   \includegraphics[width=5cm]{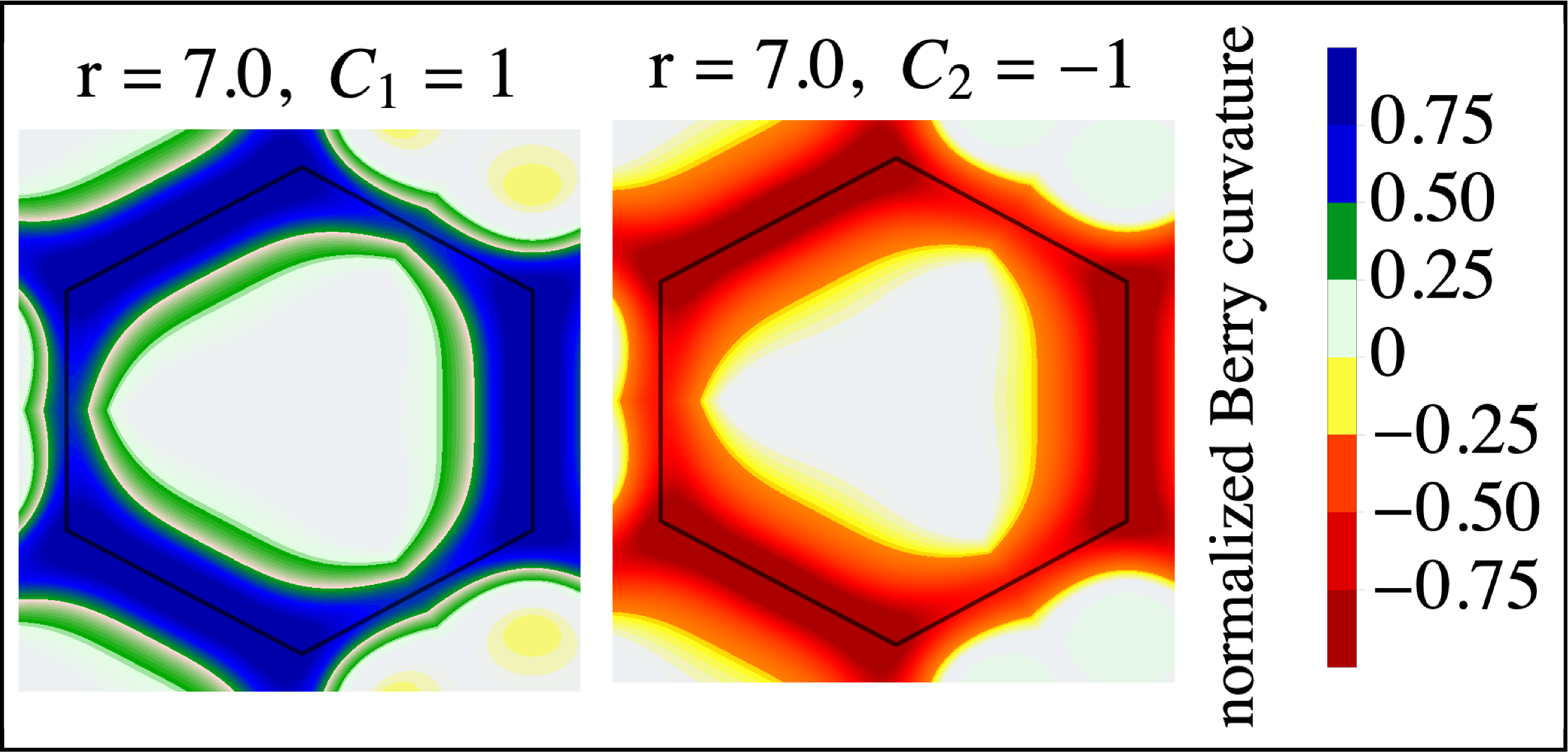}
   \caption{Berry curvature for the two lowest energy bands at strong coupling $ r=0.7$ ($t_1=1$ and $t_2=1/3$).}
          \label{}
    \vskip -0.5cm
\end{figure}

To address the strong-coupling limit $r\gg (t_1,t_2)$ analytically, we define the field operators $\psi_{\pm}=1/\sqrt{2}(c_{gA} \pm c_{hA})$ hybridizing the sublattices $A$ of the two layers and $\chi_{\pm}=1/\sqrt{2}(c_{gB} \pm c_{hB})$ hybridizing the sublattices $B$ of the two layers. Similarly to the graphene layer in Eq. (3), $c_{hA}^{\dagger}$ and $c_{hB}^{\dagger}$ represent creation operators at sublattice $A$ and $B$ in the Haldane layer. 
To show that the strong-coupling description is very general we introduce the magnetic  fields ${\bf d}_1$ and ${\bf d}_2$ associated with the two layers, that we shall rewrite in the hybridized basis. To find the effective Hamiltonian in the basis $[\psi_-, \chi_-, \psi_+, \chi_+]$, we can equivalently perform a unitary transformation on the Hamiltonian such that the Hamiltonian becomes
\begin{eqnarray}
  \tilde{ {\cal H}}({\bf k}) =
  \left({\begin{array}{cc}
- r\mathcal{I} + \frac{({\bf d}_1+{\bf d}_{2})}{2}\cdot \bm{\sigma}  & \frac{({\bf d}_1-{\bf d}_{2})}{2}\cdot \bm{\sigma} \\
 \frac{({\bf d}_1-{\bf d}_{2})}{2}\cdot \bm{\sigma}  &  r\mathcal{I} + \frac{({\bf d}_1+{\bf d}_{2})}{2}\cdot \bm{\sigma}  \\
  \end{array} } \right).
\end{eqnarray}
The energy spectrum shows two pairs of bands centered around $\pm r$ (see \cite{SM}) and described by a Haldane model with an effective magnetic field in ${\bf k}$ space which is equivalent to $({\bf d}_1+{\bf d}_2)\cdot {\bm\sigma}/2$.  The off-diagonal terms couple band pairs of different energy which do not affect the low-energy theory. For the Haldane-graphene bilayer with ${\bf d}_1={\bf d}_g$ and ${\bf d}_2={\bf d}_h$, Berry curvatures of the two lowest bands for $r\gg t_2$ are shown in Fig. 3. 

\begin{figure}[t]
   \includegraphics[width=4.5cm]{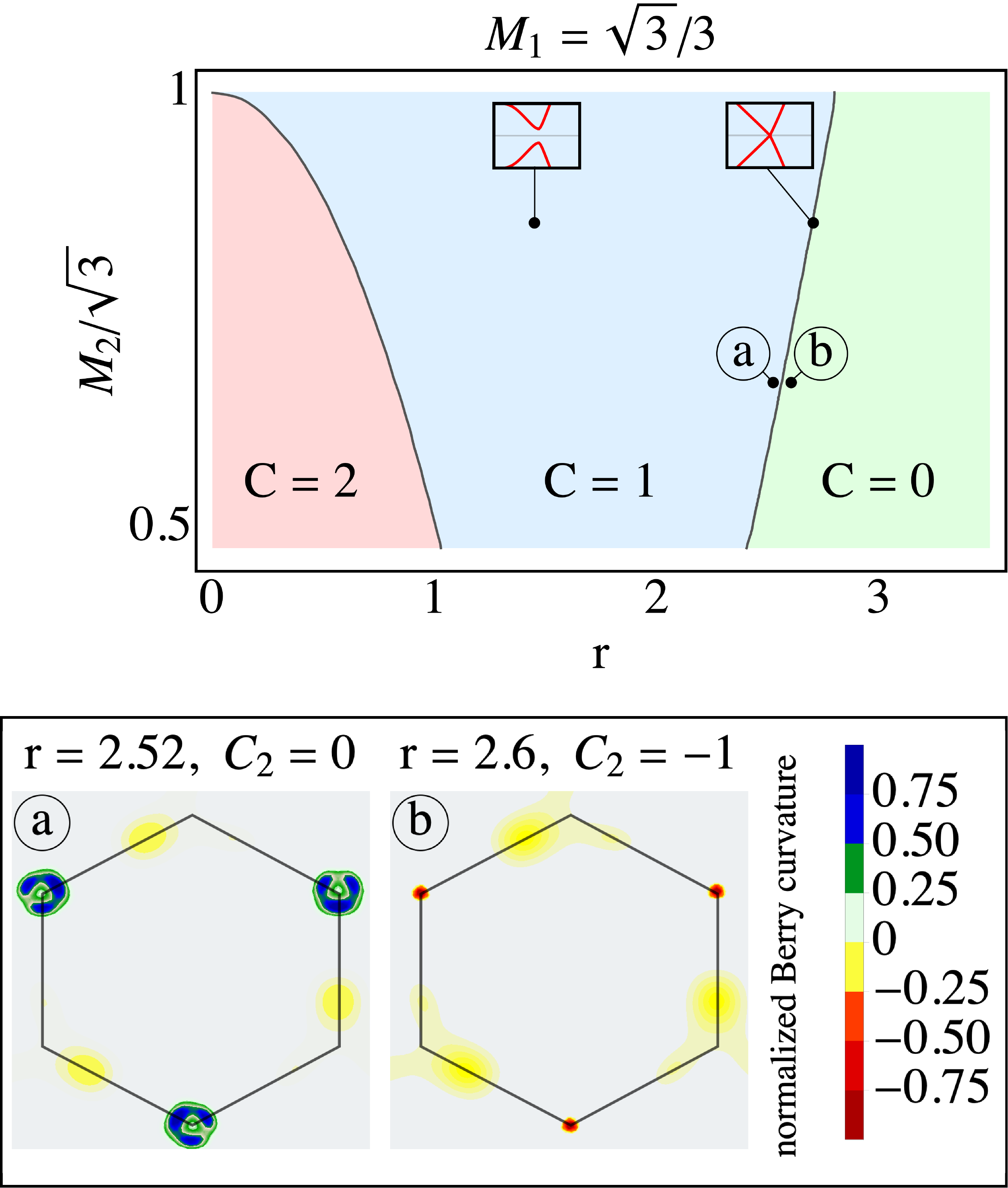}
   \caption{Numerical phase diagram for two coupled Haldane models with ${\bf d}_1= {\bf d}_2={\bf d}_h$. Evolution of the total Chern number $C$ for the two lowest bands as a function of $r$ and $M_2$ for $M_1=1/\sqrt{3}$ in units of $t_1=1$ and $t_2=1/3$. Illustration of Berry phase `jumps' at the second phase transition.}
          \label{}
    \vskip -0.5cm
\end{figure}

In Fig. 2, the two lowest ``hybrid'' bands are still described by a total Chern number zero and the bulk green region now behaves as the vacuum. In Fig. 2, we now observe two counter-propagating edge modes with equal velocities at zero energy, when suppressing the tunnel coupling at one edge in the cylinder geometry (in the grey region). By making one layer slightly larger than the other, the two edges modes now entirely connect to the Haldane bulk bands of the grey region. 

At quarter filling (implying that the particle density of the two layers satisfy $(n_g=n_h=1/4)$) only the lowest band in Fig. 2 right should be filled, and the system reveals a topological phase with Chern number 1.  The edge structure shows on average 1/2 particle in one layer moving together with 1/2 particle in the other layer. 

We now address the situation of a Haldane model in each layer with ${\bf d}_1= {\bf d}_2={\bf d}_h$ which can be realized in ultra-cold atoms through a shaking protocol, then resulting in $t_2\ll t_1$ \cite{SM}. We predict an evolution from a phase with total Chern number $C=2$ to a phase with total Chern number 0, when increasing $r$ \cite{SM}. This demonstrates that the topological proximity effect subsists in the case of two coupled insulators \cite{Frankfurt}. We now discuss the effect of Semenoff masses $M_1$ and $M_2$ in the two layers \cite{Semenoff}. This results in an additional term $(M_1+M_2)\sigma_z/2$ in Eq. (6) in the sub-space $[\psi_-,\chi_-]$ and similarly for the sub-space $[\psi_+,\chi_+]$. For asymmetric masses, we find two transitions showing a jump of Berry phase at one Dirac point only, namely the ${\bf K}'$ and then the ${\bf K}$ point, and the bands remain distinguishable in the intermediate $C=1$ region (see Fig. 4) \cite{SM}. For $M_1=M_2$, a band touching effect occurs, then suppressing the $C=1$ region \cite{SM}.

Below, we present an alternative description of the topological proximity effect on the Bloch sphere with polar angle $\theta({\bf k})$ and azimuthal angle $\bf{\phi}({\bf k})$, defining the Haldane model at $r=0$. From Stokes' theorem, for $r\neq 0$, we rewrite the Chern number of the different bands in the equatorial plane for an angle $\theta({\bf k})=\pi/2$, as 
\begin{equation}
C_j = \frac{1}{2\pi}\int_0^{2\pi} d\phi \left(\langle \psi_N^j | i\frac{\partial}{\partial \phi} | \psi_N^j\rangle - \langle \psi_S^j | i\frac{\partial}{\partial \phi} | \psi_S^j\rangle\right),
\end{equation}
where $|\psi_N^j\rangle = |\psi_j(\theta({\bf k}),\phi({\bf k}),N)\rangle$ and $|\psi_S^j\rangle = |\psi_j(\theta({\bf k}),\phi({\bf k}),S)\rangle$ are eigenstates corresponding to the band $j$ defined in the north (N) or south (S) hemisphere \cite{SM}.
Going from north to south pole is equivalent to modify the vectors ${\bf a}_i \rightarrow - {\bf a}_i$ and ${\bf b}_i \rightarrow - {\bf b}_i$ in real space (if we fix ${\bf k}$). This is also equivalent to change the role of sublattices $A$ and $B$ in each layer. 
We check that for the lowest energy band corresponding to the valence band of the Haldane model in weak coupling, then $C_1 = C_-^h=+1$.  The two lowest bands (and also the two upper energy bands) acquire opposite winding phases due to the relative phase fixing argument at the north pole when $r\neq 0$, {\it i.e.} $C_2 = C_-^g = -1$, encoding the mass inversion effects between bands. We justify the equivalence between Eqs. (2) and (7) in Ref. \cite{SM}.

To summarize, we have presented a proximity effect from a topological Chern insulator on a graphene layer.  Particle-hole processes at the interface induce a gap in the graphene layer: the two lowest filled energy bands show inverse quantized Chern numbers $+1$ and $-1$.  We have illustrated the bulk-edge correspondence in relation with the Kane-Mele model \cite{KM,Sheng}, and with general bulk-edge correspondence in the ultra strong-coupling limit.  The effective model built in ${\bf k}$-space close to the Dirac points could be realized in graphene bilayers through circularly polarized light coupling to one layer more prominently \cite{Cavalleri}. We have generalized the Berry phase jump phenomenon to bilayer Haldane models. In the Supplementary Material, we discuss implementations in cold atom and light systems thoroughly.  Interaction effects leading to Mott transition \cite{Wei,Stephan,HaldaneMott,KMboson} and fractional quantum Hall phases will be studied further \cite{Sopheak}.

We acknowledge discussions with B. A. Bernevig, I. Bloch, A. Eckardt, N. Goldman, F. Heidrich-Meisner, L. Herviou, W. Hofstetter, S. Munier, P. Paganini, S. Rachel, G. Roux, A. Subedi, L. Tarruell, J.-H. Zheng, W. Wu. 
We acknowledge funding from the Deutsche Forschungsgemeinschaft (DFG, German Research Foundation) via Research Unit FOR 2414 under project number 277974659. KLH acknowledges funding from ANR BOCA and PC from the LabeX PALM through ANR-10-LABEX-0039. This research has benefitted from discussions at CIFAR meetings in Canada. 
\\

$^{\dagger}$ {\it The two authors have contributed equally.}

$^{*}$ {\it For information, the corresponding author e-mail is: karyn.le-hur@polytechnique.edu}

\pagebreak
\onecolumngrid

\section{Definitions and Notations}
We give brief information on the definitions used in this work. The lattice vectors, see Fig. 5, are given by
\begin{equation}
{\bf u}_1 = \frac{1}{2}\left(3,\sqrt{3}\right) \hspace{1cm} {\bf u}_2 = \frac{1}{2}\left(3,-\sqrt{3}\right). \hspace{1cm} {\bf u}_3=\left(0,0\right)
\end{equation}
where we set the bond length to one. Furthermore, we denote nearest neighbor displacements by 
\begin{equation}
{\bf a}_1=\frac{1}{2}\left(1,\sqrt{3}\right)\hspace{1cm}
{\bf a}_2=\frac{1}{2}\left(1,-\sqrt{3}\right)\hspace{1cm}
{\bf a}_3=\frac{1}{2}\left(-1,0\right).
\end{equation}
The next nearest neighbour displacements in a basis of the $ {\bf a}_i $ are then expressed as $ {\bf b}_i={\bf a}_j-{\bf a}_k $, where $ \left( i,j,k \right) $ is a cyclic permutation of $ \left( 1,2,3 \right) $. However, note that using an $ {\bf a}_i $ basis does not yield a Hamiltonian in Bloch form. In practice, we therefore employ a basis given by the lattice vectors $ {\bf u}_i $ (which corresponds to a gauge transforming the Hamiltonian to the new basis) and define next nearest neighbour displacements $ {\bf b}_i $ accordingly in terms of the $ {\bf u}_i $.

Finally, the high symmetry Dirac points of the Brillouin zone are located at 
\begin{equation}
{\bf K} = \frac{2\pi}{3}\left(1,\frac{1}{\sqrt{3}}\right) \hspace{1cm} {\bf K}^{\prime} = \frac{2\pi}{3}\left(1,-\frac{1}{\sqrt{3}}\right), \label{K-points}
\end{equation}
and the lattice spacing is fixed to unity, in real space.

\begin{figure}[h]\centering
\includegraphics[width=0.7\textwidth]{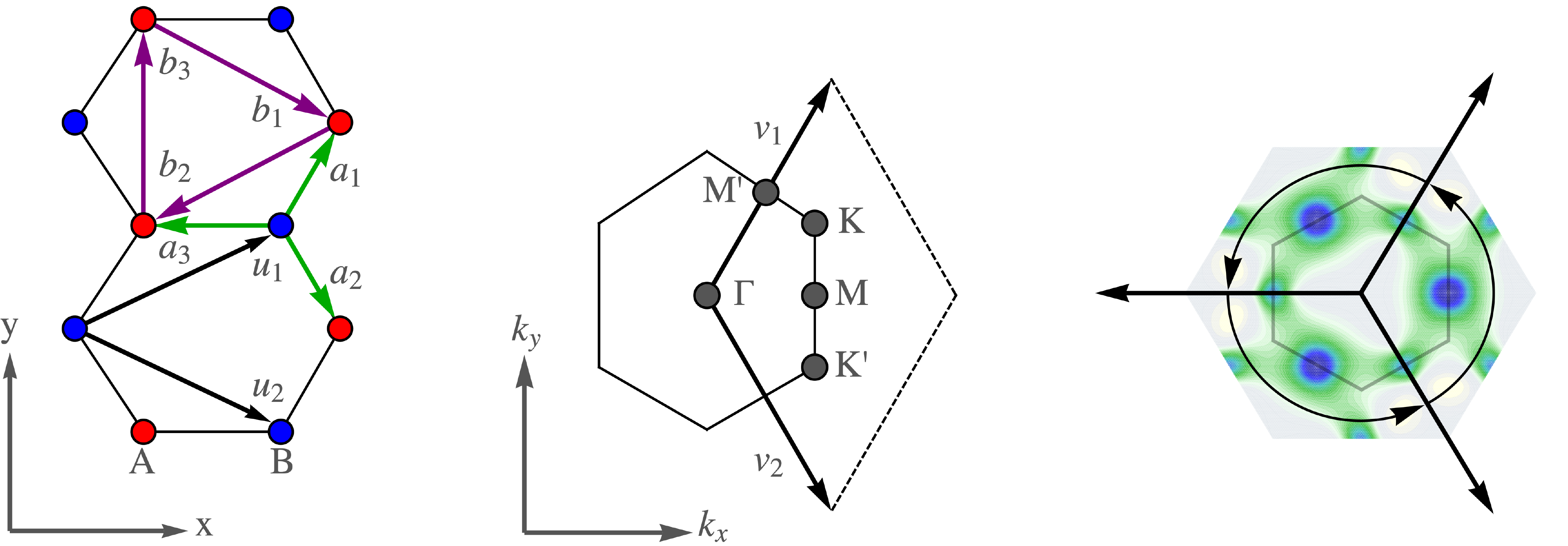}
\caption{Graphene lattice, lattice vectors, and (next) nearest neighbor displacements. Center: Brillouin zone, reciprocal lattice vectors and high symmetry points. Right: Reconstruction of the Brillouin zone for the Berry curvature plot using the $C_3$ symmetry of the Haldane model. For each plot, the result of the Berry curvature is normalised to one, i.e. each data set is divided by the maximum absolute Berry curvature value contained in the data set.}
\label{}
\end{figure}

The numerical figures of the Berry curvatures are obtained with the reconstruction of the Brillouin zone \cite{NumericsBerry}, as in Fig. 5. We observe a $C_3$ symmetry in the Berry curvature profiles of the Haldane layer, related to Fig. 1 in the Letter.  The introduction of the flux $\Phi$ in the $t_2$ term leads to a breaking of the $C_6$ symmetry, and all $M$ points are not equivalent.

\section{Berry phase shift and Eigenstates}

Upon coupling the Haldane layer to the graphene layer with finite inter-layer hopping constant $ r $, we showed that an effective diagonal term $ -r^2/(27\sin^2[\Phi]t_2^2)d_z^h\left({\bf k}\right)\sigma^z $ is induced in the graphene layer, close to the
Dirac points. In order to study the Berry phase shift of $ 2\pi $ that occurs in this scenario at one Dirac point, we investigate the low energy version of the effective graphene Hamiltonian in more detail. To this end, we first expand the term $ d^h_z\left({\bf k}\right)=-2t_2 \sin\left(\Phi\right)\sum_j\sin\left({\bf k}\cdot{\bf b}_j\right) $ around $ {\bf K}$ and  $ {\bf K}^{\prime}$ to first order for small $ {\bf q}=\left(q_x,q_y\right) $. We obtain 
\begin{equation}
d^h_z\left( \pm{\bf K}+ {\bf q} \right)  \approx \pm 3\sqrt{3} t_2  \sin\left(\Phi\right),
\end{equation}
where the positive sign corresponds to $ {\bf K} $ and the negative sign to $ {\bf K}^{\prime} $. We define $ \pm r^2\sqrt{3}/(9\sin\Phi t_2) \equiv \pm m $ which has different signs at the two Dirac points. 

Expanding the off-diagonal terms $ d_x^g\left({\bf k} \right) $ and $ d_y^g\left({\bf k} \right) $ around $ {\bf K} $ and $ {\bf K}^{\prime} $ respectively yields the corresponding low energy Hamiltonians
\begin{equation}
\mathcal{H}_{{\bf K}}\left({\bf q}\right) =
\left( \begin{array}{cc}
m &  v_F\left(-i q_x +q_y\right) \\ 
v_F\left(i q_x + q_y\right) & -m
\end{array} \right),
\hspace{1cm}
\mathcal{H}_{{\bf K}^{\prime}}\left({\bf q}\right)
=\left( \begin{array}{cc}
-m & v_F\left(-i q_x -q_y\right) \\ 
v_F\left(i q_x - q_y\right) & m
\end{array} \right),\label{H_mat}
\end{equation}
with $v_F = 3t_1/2 $ the Fermi velocity. Remember that in the case of pure graphene $ \left( m=0\right) $ diagonalization results the normalized eigenstates (in the Letter, ${\phi}({\bf q})= \phi_{\bf q}$)
\begin{equation}
\Psi^g_{\pm,{\bf K}}\left(\phi_{\bf q}\right)=\frac{1}{\sqrt{2}}
\left( \begin{array}{c}
1  \\ 
\pm e^{i\phi_{\bf q}}
\end{array} \right),
\hspace{1cm}
\Psi^g_{\pm,{\bf K}^{\prime}}\left(\phi_{\bf q}\right)=\frac{1}{\sqrt{2}}
\left( \begin{array}{c}
1 \\ 
\pm e^{-i\phi_{\bf q}} 
\end{array} \right),
\end{equation}
where $ i q_x +q_y=q\cdot e^{i\phi_{\bf q}} $ and $ q = \vert {\bf q}\vert $. Note that $ \phi_{\bf q}\rightarrow - \phi_{\bf q} $ relates the Dirac points $ {\bf K} $ and $ {\bf K}^{\prime} $ for the same energy band (denoted $ + $ or $ - $) as
\begin{equation}
\Psi^g_{\pm,{\bf K}}\left(\phi_{\bf q}\right) =
\Psi^g_{\pm,{\bf K}^{\prime}}\left(-\phi_{\bf q}\right).\label{graphene_id_wv}
\end{equation}
A straightforward diagonlization of the matrices equation (\ref{H_mat}) yields the normalized eigenstates 
\begin{align}
\tilde{\Psi}_{\pm,{\bf K}}\left(\phi_{\bf q}\right) 
&= \frac{1}{\sqrt{v_F^2 q^2 + \left( E_{\pm}\left({\bf q}\right) - m  \right)^2}}\left( \begin{array}{c}
v_F q \\ 
e^{i\phi_{\bf q}}\left(  E_{\pm}\left({\bf q}\right) - m   \right)
\end{array}  \right), \label{psi1}\\
\tilde{\Psi}_{\pm,{\bf K}^{\prime}}\left(\phi_{\bf q}\right) 
&= \frac{1}{\sqrt{v_F^2 q^2 + \left( E_{\pm}\left({\bf q}\right) + m  \right)^2}}\left( \begin{array}{c}
v_F q \\ 
-e^{-i\phi_{\bf q}}\left(  E_{\pm}\left({\bf q}\right) + m   \right)
\end{array}  \right), \label{psi2}
\end{align}
where the corresponding energy eigenvalues are $ E_{\pm}\left({\bf q}\right)=\pm\sqrt{v_F^2 q^2 + m^2} $.

The wavefunctions $ \tilde{\Psi}_{-,{\bf K}} $ and $ \tilde{\Psi}_{+,{\bf K}^{\prime}} $ are well defined in the limit $ {\bf q}\rightarrow 0 $. Crucially however, $ \tilde{\Psi}_{+,{\bf K}} $ and  $ \tilde{\Psi}_{-,{\bf K}^{\prime}} $ become singular as $ E_{\pm}(q) \xrightarrow{\textbf{q}\rightarrow 0} \pm m $. Hence, the wavefunction  $ \tilde{\Psi}_{+,{\bf \cdot}} $ has a singularity in $ {\bf K} $ and the wavefunction $ \tilde{\Psi}_{-,{\bf \cdot}} $ has a singularity in $ {\bf K}^{\prime} $. The emergence of these singularities in the wavefunctions signals that the coupling to the Haldane layer induced some non-trivial topology in the graphene layer. Non-trivial topolgy arises when no global phase convention can be determined in the Brillouin zone causing the wavefunction to develop singularities \citep{Kohmoto}. However, the singularities can be avoided.

First, note that the wavefunctions equations (\ref{psi1}) and (\ref{psi2}) fulfill the following identities
\begin{equation}
\tilde{\Psi}_{\pm,{\bf K}} \left(\phi_{\bf q}\right) =
\tilde{\Psi}_{\mp,{\bf K}^{\prime}} \left(-\phi_{\bf q}\right). 
\end{equation}
Hence, contrary to equation (\ref{graphene_id_wv}) for pure graphene and in agreement with the reasoning in the Letter, substituting $ \phi_{\bf q}\rightarrow - \phi_{\bf q} $ relates the wavefunction of the positive (negative) energy band at $ {\bf K} $ with the wavefunction of the negative (positive) energy band at $ {\bf K}^{\prime} $. In line with this, we can conclude that in the pure graphene limit $ m\rightarrow 0 $ we regain:
\begin{equation}
\tilde{\Psi}_{\pm,{\bf K}} \left(\phi_{\bf q}\right) \xrightarrow{m\rightarrow 0} \Psi^g_{\pm,{\bf K}} \left(\phi_{\bf q}\right),\hspace{1cm}
\tilde{\Psi}_{\pm,{\bf K}^{\prime}} \left(\phi_{\bf q}\right) \xrightarrow{m\rightarrow 0} \Psi^g_{\mp,{\bf K}^{\prime}} \left(\phi_{\bf q}\right).
\end{equation}
We now follow the method outlined in \citep{Kohmoto} and divide the Brillouin zone into two sectors $ \mathcal{S} $ and $ \mathcal{S}^{\prime} $, where sector $ \mathcal{S} $ contains $ {\bf K} $ and sector $ \mathcal{S}^{\prime} $ contains $ {\bf K}^{\prime} $. We focus on the negative energy band. $ \tilde{\Psi}_{-,{\bf \cdot}} $ is well defined in $ \mathcal{S} $, but becomes singular in $ \mathcal{S}^{\prime} $. As $ \tilde{\Psi}_{-,{\bf K}} \left(\phi_{\bf q}\right) =\tilde{\Psi}_{+,{\bf K}^{\prime}} \left(-\phi_{\bf q}\right)$ we can identify $ \tilde{\Psi}_{+,{\bf K}^{\prime}} \left(-\phi_{\bf q}\right) $ as a well defined wavefunction in $ \mathcal{S}^{\prime} $ of the negative energy band. This indicates that for $ {\bf K}^{\prime} $ the positive and negative energy bands exchanged their nature upon coupling the graphene and Haldane layers. In fact, it is suggestive to redefine the wavefunctions as follows where the new wavefunction $ \Psi_{\pm,{\bf \cdot}} $ is valid in each respective sector and energy band
\begin{align*}
\Psi_{+,{\bf K}}\left(\phi_{\bf q}\right) &\equiv 
\tilde{\Psi}_{-,{\bf K}}\left(-\phi_{\bf q}\right),
&\Psi_{+,{\bf K}^{\prime}}\left(\phi_{\bf q}\right) &\equiv 
\tilde{\Psi}_{+,{\bf K}^{\prime}}\left(\phi_{\bf q}\right),\\
\Psi_{-,{\bf K}}\left(\phi_{\bf q}\right) &\equiv 
\tilde{\Psi}_{-,{\bf K}}\left(\phi_{\bf q}\right),
&\Psi_{-,{\bf K}^{\prime}}\left(\phi_{\bf q}\right) &\equiv 
\tilde{\Psi}_{+,{\bf K}^{\prime}}\left(-\phi_{\bf q}\right).
\end{align*} 
Writing these wavefunctions explicitly yields
\begin{equation}
\Psi_{\pm,{\bf K}/{\bf K}^{\prime}}\left(\phi_{\bf q}\right) 
= \frac{1}{\sqrt{v_F^2 q^2 + \left(E_{\pm}\left({\bf q}\right) \pm m  \right)^2}}\left( \begin{array}{c}
v_F q\\
\mp e^{\mp i\phi_{\bf q}}\left( E_{\pm}\left({\bf q}\right) \pm m  \right) 

\end{array}  \right). \label{psi_new}\\
\end{equation}
This ``patching'' of wavefunctions in sectors is allowed as long as the wavefunctions are connected by a smooth gauge transformation at the boundary between the sectors \citep{Kohmoto}. Note that $ \Psi_{\pm,{\bf \cdot}} $ is of the same form in $ \mathcal{S} $ and $ \mathcal{S}^{\prime} $. Therefore, the gauge transition function between $ \mathcal{S}$ and $ \mathcal{S}^{\prime} $ is the identity. This means that $ {\bf K} $ and $ {\bf K}^{\prime} $ have the same Berry phase. 

Thus, by imposing finite coupling $ r $ the wavefunction becomes singular in one sector. The singularity can be lifted upon exchanging positive and negative energy bands in this sector. Therefore, the Berry phase jumps by $ 2\pi $ at only one Dirac point. This proof gives another justification to equation (5) in the Letter. 

\section{Topological Invariant and Eigenstates' Evolution}

Here, we suggest a formulation of the topological invariant in the equatorial plane.  We fix $d_x^h=d_x^g=d_x$ and $d_y^h=d_y^g=d_y$. 
In the equatorial plane, $d_z^h = d_z^g = 0$, and assuming the same nearest-neighbor hopping amplitude in the two layers, the Hamiltonian takes the form:
\begin{eqnarray}
   {\cal H}_{equator} =
  \left({\begin{array}{cccc}
0  & d_x - i d_y & r & 0 \\
 d_x+id_y & 0 & 0& r \\
r & 0 & 0& d_x-i d_y \\
0 & r & d_x + i d_y & 0 \\
  \end{array} } \right)
\end{eqnarray}
The four eigenstates take the form (with energies $E_1<E_2<E_3<E_4$) 
\begin{eqnarray}
|\phi_1\rangle =
  \left({\begin{array}{c}
- \frac{d_x - i d_y}{\sqrt{d_x^2+d_y^2}} \\
1  \\
\frac{d_x-i d_y}{\sqrt{d_x^2+d_y^2} }\\
-1 \\
  \end{array}} \right)
\hskip 1cm  \hbox{with energy} \hskip 1cm
  E_1 = -r - \sqrt{d_x^2+d_y^2}.
\end{eqnarray}
\begin{eqnarray}
|\phi_2\rangle =
  \left({\begin{array}{c}
- \frac{d_x - i d_y}{\sqrt{d_x^2+d_y^2}} \\
1  \\
- \frac{d_x-i d_y}{\sqrt{d_x^2+d_y^2} }\\
1 \\
  \end{array}} \right)
\hskip 1cm  \hbox{with energy} \hskip 1cm
  E_2 = r - \sqrt{d_x^2+d_y^2}.
\end{eqnarray}
\begin{eqnarray}
|\phi_3\rangle =
  \left({\begin{array}{c}
  1 \\
 \frac{d_x + i d_y}{\sqrt{d_x^2+d_y^2}} \\
- 1  \\
- \frac{d_x-i d_y}{\sqrt{d_x^2+d_y^2} }\\
  \end{array}} \right)
\hskip 1cm  \hbox{with energy} \hskip 1cm
  E_3 = -r + \sqrt{d_x^2+d_y^2}.
\end{eqnarray}
\begin{eqnarray}
|\phi_4\rangle =
  \left({\begin{array}{c}
  1 \\
 \frac{d_x + i d_y}{\sqrt{d_x^2+d_y^2}} \\
 1  \\
\frac{d_x-i d_y}{\sqrt{d_x^2+d_y^2} }\\
  \end{array}} \right)
\hskip 1cm  \hbox{with energy} \hskip 1cm
  E_4 = r + \sqrt{d_x^2+d_y^2}.
\end{eqnarray}
Now, we fix precisely the gauge between these different eigenstates by taking a path at fixed angle $\phi$ and going to the north pole where $d_x=d_y=0$ and the coupled layers are described through the $r$ coupling and
the $d_z^h$ field for the Haldane model. This is equivalent to go exactly to a Dirac point in the first Brillouin zone. 
We then find the four eigen-energies
\begin{eqnarray}
E_{1,3} &=& \frac{1}{2}\left(-d_z^h \mp \sqrt{ (d_z^h)^2 + 4r^2}\right) \\ \nonumber
E_{2,4} &=& \frac{1}{2}\left(d_z^h \mp \sqrt{ (d_z^h)^2 + 4r^2}\right).
\end{eqnarray}
The four associated eigenstates at the north pole take the form
\begin{eqnarray}
|\phi_1\rangle ' =
  \left({\begin{array}{c}
0 \\
0  \\
0 \\
1 \\
  \end{array}} \right)
  \hskip 0.5cm
  |\phi_2\rangle ' =
  \left({\begin{array}{c}
0 \\
1  \\
0 \\
0 \\
  \end{array}} \right)
    \hskip 0.5cm
   |\phi_3\rangle ' =
  \left({\begin{array}{c}
1 \\
0  \\
0 \\
0 \\
  \end{array}} \right) 
    \hskip 0.5cm
   |\phi_4\rangle ' =
  \left({\begin{array}{c}
0 \\
0  \\
1 \\
0 \\
  \end{array}} \right).
  \end{eqnarray}
  These states could be redefined modulo a general global phase, but independent of $\phi$ which is not well defined at the north pole, 
  In agreement with the path-integral argument in the Letter and the weak-coupling argument,  in the limit of small $r$ we find the four eigen-energies (and assume here $d_z^h>0$):
  \begin{eqnarray}
  E_{1,4} &\approx& \mp d_z^h \nonumber \\
  E_{2,3} &\approx& \mp \frac{r^2}{d_z^h}. 
  \end{eqnarray}
  The mass inversion phenomenon found when changing ${\bf K}$ into ${\bf K}'$ occurs when going to the south pole, where $\theta\rightarrow \pi-\theta$, therefore $d_z^h \rightarrow - d_z^h$.
  
Decreasing progressively the angle $\theta$ from the north pole until $d_z^h \sim r$, to build a precise correspondence between the $|\phi_i\rangle$ states at the equator and the $|\phi_i\rangle'$ states from the north pole, then we identify precisely the dependence on r of the different eigen-energies.  We then infer the following correspondence $|1\rangle ' \leftrightarrow |1\rangle$, $|3\rangle ' \leftrightarrow |2\rangle$, $|2\rangle ' \leftrightarrow |3\rangle$, $|4\rangle ' \leftrightarrow |4\rangle$ for
$d_z^h\sim r$.  We can now precisely fix the relative gauge (in terms of the $\phi$ variable) between different eigenstates in the equatorial plane in agreement with Eqs. (26).  

Then, we identify:
\begin{eqnarray}
|\phi_1\rangle  =
  \left({\begin{array}{c}
- e^{-i \phi} \\
1  \\
e^{-i \phi} \\
-1 \\
  \end{array}} \right)
  \hskip 0.5cm
  |\phi_2\rangle  =
  \left({\begin{array}{c}
1 \\
- e^{i \phi}  \\
1 \\
- e^{i \phi} \\
  \end{array}} \right)
    \hskip 0.5cm
   |\phi_3\rangle  =
  \left({\begin{array}{c}
e^{-i\phi} \\
1  \\
- e^{-i \phi} \\
-1 \\
  \end{array}} \right) 
    \hskip 0.5cm
   |\phi_4\rangle  =
  \left({\begin{array}{c}
1 \\
e^{i\phi}  \\
1 \\
e^{i \phi} \\
  \end{array}} \right).
  \end{eqnarray}
  Applying the protocol of the Letter, then we can define equivalently eigenstates with angles defined at the south pole, by changing $\theta\rightarrow \pi-\theta$ and $\phi\rightarrow -\phi$ and by swapping the role of sublattice $A$ and $B$ in each layer (to match with the changes in the $d_x$, $d_y$ and $d_z^h$ fields), implying
  \begin{eqnarray}
  |\phi_1\rangle^S  =
  \left({\begin{array}{c}
1 \\
-e^{i \phi}  \\
-1 \\
e^{i \phi} \\
  \end{array}} \right) = - e^{i\phi} |\phi_1\rangle^N
    \hskip 0.5cm
  |\phi_2\rangle^S  =
  \left({\begin{array}{c}
-e^{-i\phi} \\
1  \\
-e^{-i\phi} \\
1 \\
  \end{array}} \right) = -e^{-i\phi} |\phi_2\rangle^N
  \end{eqnarray}
  \begin{eqnarray}
   |\phi_3\rangle^S  =
  \left({\begin{array}{c}
1 \\
e^{i\phi}  \\
-1 \\
-e^{i\phi} \\
  \end{array}} \right) = e^{i\phi} |\phi_3\rangle^N
    \hskip 0.5cm
   |\phi_4\rangle^S =
  \left({\begin{array}{c}
e^{-i\phi} \\
1 \\
e^{-i\phi} \\
1 \\
  \end{array}} \right) = e^{-i\phi}|\phi_4\rangle^N.
  \end{eqnarray}
  
  The important point is that due to the identification between the states $|\phi_2\rangle' = |\phi_3\rangle$ and $|\phi_3\rangle' = |\phi_2\rangle$, the bands 1 and 3 become defined by the same integer number
  \begin{eqnarray}
  C_1' = C_3' = \frac{1}{2\pi}\int_0^{2\pi} d\phi \left( {^N\langle} \phi_j | i\frac{\partial}{\partial \phi} | \phi_j\rangle^N - {^S\langle}\phi_j | i\frac{\partial}{\partial \phi} | \phi_j\rangle^S\right) = +1.
  \end{eqnarray}
  In a similar manner, we identify $C_2' = C_4' = -1$. This number defined on the circle in the equatorial plane then contains information on the topological nature of the different bands.  It is important to notice
  that the transformation $\theta\rightarrow \pi-\theta$ is equivalent to change the $d_z^h\rightarrow -d_z^h$ in the Haldane model, therefore modifying $\theta\rightarrow \pi-\theta$ and $\phi\rightarrow -\phi$ reproduces the mass inversion effects around the two Dirac points. At weak-coupling, the lowest band can be identified as the lowest Haldane band and then we identify $C_-^h =C_1' = +1$. Similarly, the band $2$ can be identified as the graphene lowest band and we recover $C_2' = C_-^g=-1$. 
  \\
  
  In the Letter, we define $|\phi_j\rangle^N=|\psi_N^j\rangle = |\psi_j(\theta({\bf k}),\phi({\bf k}),N)\rangle$ and $|\phi_j\rangle^S=|\psi_S^j\rangle = |\psi_j(\theta({\bf k}),\phi({\bf k}),S)\rangle$. 
  \\
  
To justify why $C_j'$ can be identified as the topological Chern number of each band, now we apply the Stokes' theorem. Let us start with the limit $r=0$. For the Haldane model, one can decompose the sphere onto two hemispheres (north and south), and due to different orientations of surfaces on the two hemispheres, then write the Chern number of the upper Haldane band as
  \begin{equation}
  C_+^h = \frac{1}{2\pi}\int_{north.hemisphere} d{\bf n}\cdot
  \left( \bm{\nabla}\times {\bf A}_N - \bm{\nabla}\times {\bf A}_S\right),
  \end{equation}
  with the Berry connections ${\bf A}_N = {\bf A}(\phi,\theta)$ and ${\bf A}_S={\bf A}(-\phi,\pi-\theta)$, and ${\bf n}$ is a vector defining the surface of the north hemisphere.  Applying the Stokes' theorem, then we also identify
    \begin{equation}
 C'_4 = C_+^h = \frac{1}{2\pi} \int_0^{2\pi} d\phi (A(\phi,\theta=\pi/2)  - A(-\phi,\theta=\pi/2)) = -1.
  \end{equation}
  We identify for this band, with the gauge fixing arguments at the north pole in Eq. (26) for the state $|\phi_4\rangle'$, $A(\phi,\theta=\pi/2)=-1/2$ and $A(-\phi,\theta=\pi/2))=+1/2$. More precisely, for the upper band of the Haldane model at $r=0$
  \begin{equation}
  {\bf A}= -\frac{\sin^2 \theta_k}{2}\bm{\nabla} \phi_{\bf k}.
  \end{equation}
    The key point now is that for $r\neq 0$, the north pole argument of Eq. (27) encodes the occurrence of the topological proximity effect, therefore one can now apply the Stokes' theorem for both bands, graphene and Haldane bands. The graphene bands show a mass inversion compared to those of the Haldane bands which is then equivalent to invert the definitions of the orientation of the surface for the graphene layer compared to the Haldane layer, ${\bf n}\rightarrow -{\bf n}$, then leading to $C'_3 = + 1=C_+^g$ from Stokes' theorem. 
 
 To summarize, for $r\neq 0$, we then conclude that $C'_i$ is equivalent to the Chern number of a band as a result of the Stokes' theorem since  $d_z^g$ now also becomes non-zero. Defining the Berry connection (Berry potential) as 
 $\langle \phi_j | i\frac{\partial}{\partial \phi} | \phi_j\rangle$, then for all values of $r$, we conclude that from Stokes' theorem, $C'_j$ defines in Eq. (31) is equivalent to the Chern number in Eq. (2).
    \\
    
  For completeness, we show the evolution of the energy eigenstates as a function of $r$. One identifies clearly the passage from the weak-coupling limit $(r<t_2)$ where the low-energy physics occurs at the Dirac points of graphene to the strong-coupling limit $(r\gg t_2)$ giving rise to (almost) flat bands.
    
   \begin{figure}[h]
   \includegraphics[width=12cm]{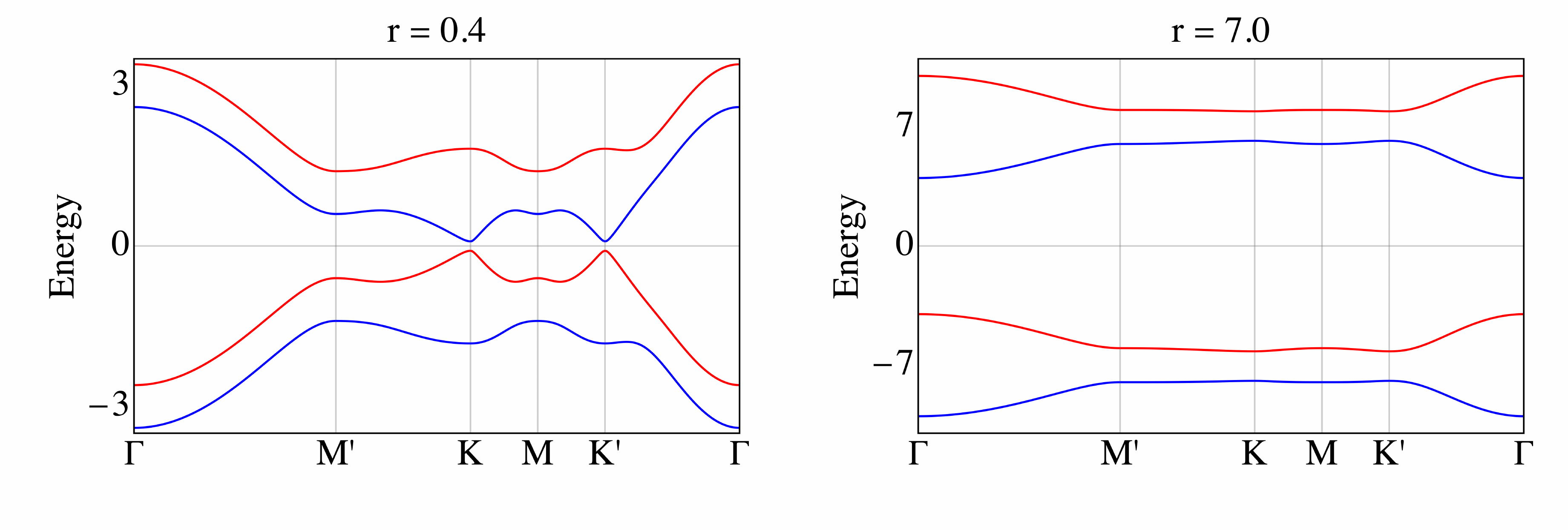}
   \caption{Evolution of the four energy eigenstates for two values of the tunnel coupling element $r$ between layers with $t_1=1$ and $t_2=1/3$. The bands are now coloured according to their Chern number. Blue means Chern number $+1$ and red Chern number $-1$.
   The $\Gamma$, $K$, $K'$, $M$ and $M'$ points are defined in Fig. 5.}
          \label{}
    \vskip -0.5cm
\end{figure}

  \section{Realizations and Predictions}
  
  \subsection{Floquet protocol in optical lattices}
  
  In optical lattices, one can apply a time-dependent force ${\bf F}(t)=-m \ddot{\bf r}_{lat}(t)$ corresponding to a periodic shaking protocol of the lattice. The Hamiltonian then becomes transformed into
  \begin{equation}
  {\cal H}_{lat} = {\cal H}_0 +\sum_i ({\bf F}(t)\cdot {\bf r}_i) c_i^{\dagger} c_i.
  \end{equation}
 Here, $c_i$ corresponds to an atom at site $i$ with mass $m$ on a honeycomb optical lattice and ${\cal H}_0$ corresponds to the Hamiltonian of graphene with nearest-neighbor tunneling coupling. The additional momentum can be absorbed by going to the reference frame $-{\bf q}_{lat}=-m\dot{\bf r}_{lat}(t)$. In this frame, the tight-binding Hamiltonian
 corresponding to nearest-neighbor tunneling becomes modified as
 \begin{equation}
 {\cal H}_{lat} ' = \sum_{\langle i; j\rangle} e^{i {\bf q}_{lat}\cdot{\bf r}_{ij}} t_{ij} c_i^{\dagger} c_j.
 \end{equation}
 In the case of a periodically driven system, where ${\cal H}_{lat}'$ and therefore ${\bf r}_{lat}(T)$ are periodic functions in time, one can then apply the Floquet theory, where an effective Hamiltonian is obtained from the unitary time-evolution operator $U(T,0)$ over one period $T$ of driving, such that 
 \begin{equation}
 \frac{i\hbar}{T}\log(U(T,0))={\cal H}_{eff}.
 \end{equation}
 Using the shaking procedure for the honeycomb optical lattice, one can then realize an effective Hamiltonian in the wave-vector space \cite{Hamburg}
 \begin{eqnarray}
 {\cal H}({\bf k}) = 
   \left({\begin{array}{cc}
 M+\sum_i 2t_{AA}\cos({\bf k}\cdot {\bf b}_i)  & \hskip 0.5cm \sum_i 2 t_{AB}e^{-i{\bf k}\cdot{\bf a}_i} \\
\sum_i 2 t_{AB}e^{i{\bf k}\cdot{\bf a}_i}  &  -M+\sum_i 2t_{BB}\cos({\bf k}\cdot {\bf b}_i)  \\
  \end{array} } \right),
 \end{eqnarray}
 acting on the Hilbert space of sublattices $A$ and $B$. The offset $M$ between A and B sites corresponds to the Semenoff mass \cite{Semenoff}. The hopping term $t_{AB}$ contributes to the nearest-neighbor graphene term $t_1$, whereas 
 $t_{AA}$ and $t_{BB}$ generate the $t_2$ terms in the Haldane model.  To realize the topological phase of the Haldane model, the key point is to use phase factors in the time-modulation of the lattice such that $t_{AA}=-t_{BB}$ and $t_{AA}=|t_2|e^{i\Phi}$, where the phase $\Phi$ corresponds to the phase introduced in Eq. (1) of the Letter. 
 
The goal is to build, for instance, two graphene optical lattices.  Then, one could apply the same time-dependent force or Floquet modulation on the two layers, as described above, to implement the same parameters $t_1$ and $t_2$ in the two layers. 
In the next step, laser assisted tunneling generates the coupling of atoms of one layer to those of the other layer, such that the $r$ tunnel coupling would couple sublattices $A$ of the two layers on the one hand and sublattices $B$ of the two layers on the other hand. Another possibility would be to use one optical lattice and two species (of synthetic dimensions). We show below how to realize the phase diagram of Fig. 4 in the Letter, through two different off-set conditions $M_1$ and $M_2$ in the two layers.
 
 Because the lattice shaking will act globally on both layers, the Haldane layer and the graphene layer have to be distinguished by the offset M in the static system before shaking. Using the different resonance condition, one can then engineer one layer in the C=1 regime and the other layer in the C=0 regime. A true graphene layer with zero offset in the Floquet system can only be approximated with circular shaking. The layer-dependent offset could be realized by using an artificial dimension in the direction of the layer, {\it i.e.} by realizing the layers as two internal Zeeman states. Using spin dependent lattice \cite{Hamburggraphene}, the two layers will naturally have different offsets, possibly even of different sign. Extending state tomography schemes to such a system would yield separate signals for the two layer, when adding a Stern-Gerlach separation during the time-of-flight expansion.
 
  \begin{figure}[h]
   \includegraphics[width=13.5cm]{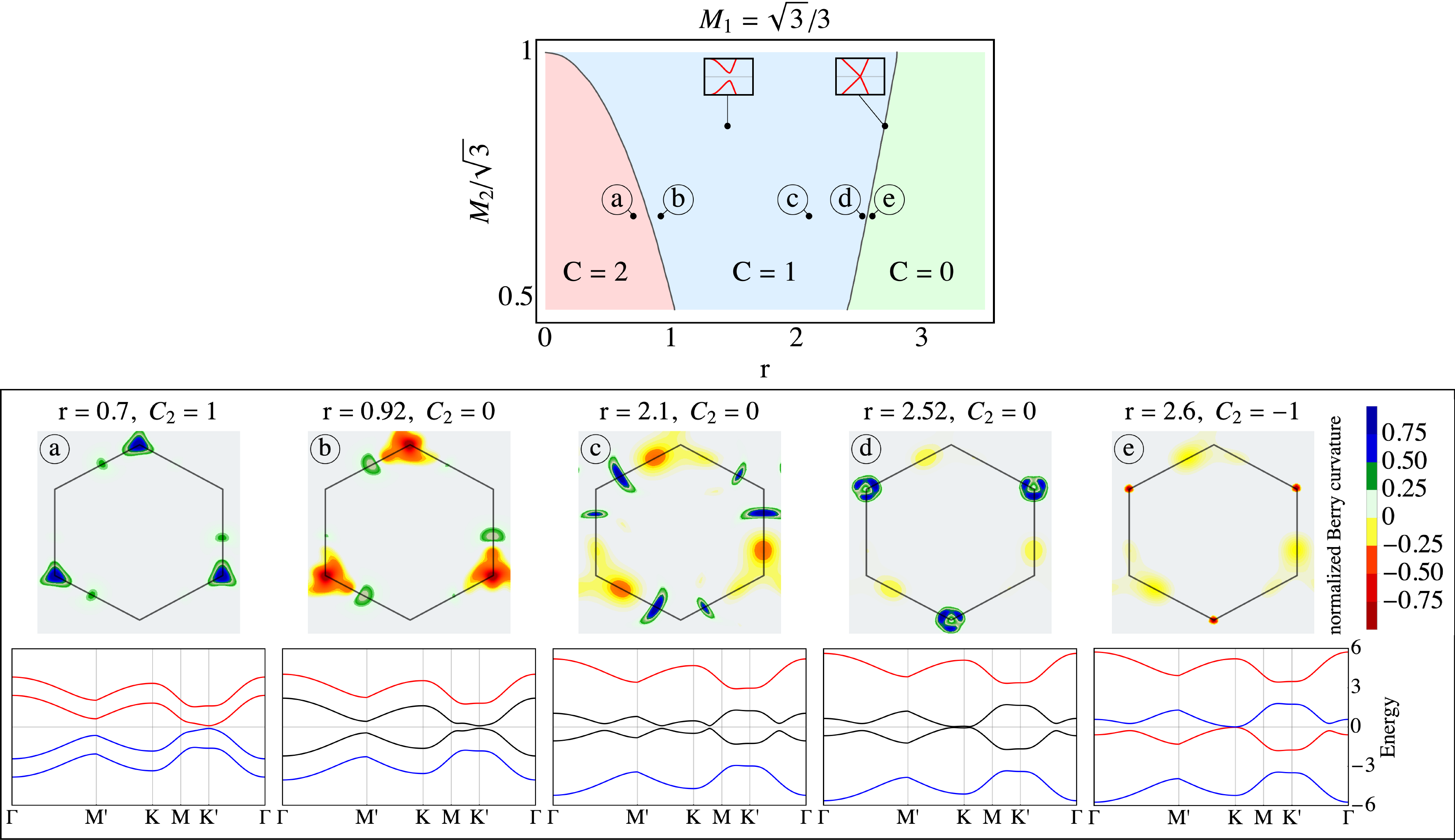}
   \caption{Upper: Numerical Phase diagram for two Haldane layers at half-filling ($t_1=1$, $t_2=1/3$, $\Phi=\pi/2$) with different Semenoff masses, where $M_1=\sqrt{3}/3$ is fixed. Lower: Berry curvatures and band structures for increasing coupling $r$ along $M_2=2\sqrt{3}/3$. At the first (second) phase transition, the bands touch at the ${\bf K}^{\prime}$ (${\bf K}$) point and the Berry curvature flips sign in its vicinity. In the band structures, the colors refer to the Chern number of the bands: Blue means Chern number $+1$, black Chern number $0$, and red Chern number $-1$. The properties of the Bloch bands could be checked with fermions and bosons, at a single particle level.}
          \label{}
\end{figure}
  
  \subsection{Specific Implementations with Ultra-cold atoms}
 
To observe the jump of Berry phase as described in Fig. 4 of the Letter, we suggest to start with two different off-sets $M_1$ and $M_2$ in the two Haldane layers. In both layers, one starts with $M_1$ and $M_2$ smaller than $|d_z({\bf k} \sim \pm {\bf K})|=3\sqrt{3}t_2\sin\Phi$. In the absence of coupling between the two layers, then the two lowest Bloch bands are described by a Chern number $+1$, producing a phase with total Chern number $C=2$. We start with both layers in the topological phase of
the Haldane model. 
 
Assuming unequal masses $M_1$ and $M_2$, we observe two phase transitions by switching on the coupling parameter $r$. At the two transitions, 
 we report a jump of Berry phase at one Dirac point only by analogy to the situation of the Haldane-graphene layers' situation at small $r$.  If we start with $M_2>M_1$ (as in Fig. 7), the gap for the bands $2$ at the ${\bf K}'$ point is (much) smaller than the gap separating the upper and lower bands $4$ and $1$ and therefore second-order processes or particle-hole pair virtual processes through these bands can still affect the gap of band $2$, which then explains the gap closing at the ${\bf K}'$ point at the first transition. We qualitatively predict by reproducing the arguments of Eq. (3) in the Letter that the gap would close at the ${\bf K}'$ point roughly when
 \begin{equation}
 3\sqrt{3}t_2\sin\Phi - M_2 - r^2/(3\sqrt{3}\sin \Phi t_2-M_1) \approx 0. 
 \end{equation}
 We check numerically that this equation reproduces the features of the first transition line.  At large $r$, the total Chern number of the two lowest bands must be zero in agreement with the theory (page 4 of the Letter). We show the band structure and Berry curvature evolve as a function of $r$, in particular for the intermediate region with $C=1$, where the gap at the ${\bf K}$ point progressively diminishes whereas the gap at the ${\bf K}'$ point now stays finite. When the gap closes at the ${\bf K}$ point, then we again observe a sign change of the Berry curvature at this point, then producing the entrance towards the $C=0$ phase.  Essentially, to enter the $C=0$ phase, the band $2$ must flip its Chern number to $C_2=-1$ then closing the gap at the ${\bf K}$ point. 
 
 For equal masses $M_1=M_2$,  a band crossing effect occurs in the intermediate region for $r\sim 0.9$, therefore the total Chern number of the two lowest bands progressively change from $C=2$ to $C=0$, as described in the figure below.
 The two phase transitions then do not occur for this case, and there is a band inversion between band $2$ and band $3$ when $r=3t_1$ for $t_2=1/3$.  But, as soon as $M_1\neq M_2$,  the system tends to restore the $C=1$ region as well as the two transitions associated with the changes in the Berry curvatures at the Dirac points. 
 
 We shall mention that band crossing effects also occur in the graphene-Haldane model in the intermediate window $r\sim 1-3t_1$, but the total Chern number of the two lowest bands remain equal to zero for all values of $r$, in 
 agreement with numerical results and strong-coupling arguments. 
 
   \begin{figure}[h]
   \vskip -0.2cm
   \includegraphics[width=17cm]{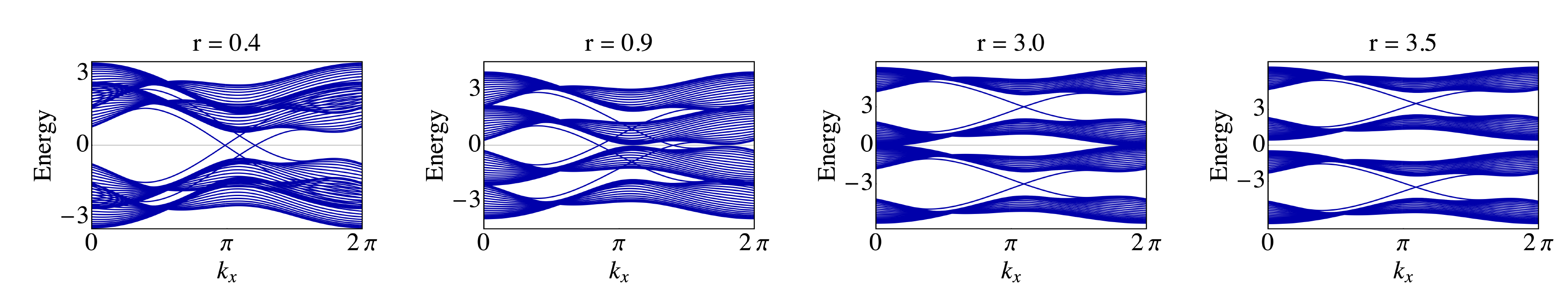}
   \caption{Evolution of edge spectra in the case of two Haldane layers, with masses $M_1 = M_2 = \sqrt{3} / 3$. Again, we fix $t_1=1$, $t_2=1/3$, $\Phi=\pi/2$.}
          \label{}
\end{figure}
  
\subsection{Berry curvature analysis}

To reconstruct the Berry curvature of the bands (quasi-bands in the Floquet basis), one can apply the protocol of Refs. \cite{Hamburg,Eckardt} and project onto flat bands. To reconstruct the Berry curvature and measure momentum space density after time-of-flight, it is required to apply a quench producing an additional $\Delta_{AB}$ off-set for the two sublattices of a given layer, described on page 2 in the Letter.  This quench has a similar effect as a rotation perpendicular to the $z$-axis, producing $\phi(t)\rightarrow \phi(t)+\Delta_{AB} t$ with $\Delta_{AB}\gg (M_1,M_2)$. After the time of flight (TOF), this yields an oscillation of the density at each momentum.  Absorption images after TOF reveal a momentum distribution where the $A$, $B$ populations, corresponding to the lowest band(s), are mapped onto the first Brillouin zone. Here, we assume that gaps are (much) larger than temperature effects. This procedure allows for a precise measurement of the angle $\phi$, and therefore of $\theta$ for the two lowest bands, through the formula of occupancies of the bands (page 2 in the Letter). Performing measurements of the density profiles in momentum space for the two layers, {\it i.e.} measuring $n_1({\bf k})$ and $n_2({\bf k})$, one could rebuild informations on the different bands through the functions of $n_1({\bf k})+n_2({\bf k})$ and $n_1({\bf k})-n_2({\bf k})$. For the Haldane bilayer system, the two lowest bands should be described by similar angles $\theta$ at small $r$, whereas at large $r$ the second lowest band is subject to a mass inversion phenomenon which is equivalent to $\theta\rightarrow \pi-\theta$ and $\phi\rightarrow -\phi$, then changing the direction of the $d_z$ component, in agreement with Eq. (7) in the Letter. To obtain additional information on the topology, one could couple the two lowest bands through circular shaking (which mimicks the effect of circularly polarized light on real materials) \cite{Nathan}. 
\\

To measure directly jumps of Berry phases at the two transitions, one could also use Wilson line techniques \cite{Munich} which measure changes in the band populations under the influence of an external force $\tilde{{\bf F}}$ and transports atoms in the reciprocal space.  One could for instance measure directly the Berry phases accumulated when encircling the Dirac points ${\bf K}$ and ${\bf K}'$ of band $2$. This is also related to the alternative definition of the topological invariant suggested in Eq. (7) on the sphere. Indeed, going from north to south poles is also equivalent to change ${\bf K}$ in ${\bf K}'$, and the measure $d\phi$ in Eq. (7) means that we now consider a similar closed path around each Dirac point.  The quasi-wavevector changes ${\bf k}(t)\rightarrow {\bf k}(0)+\tilde{{\bf F}}t/\hbar$.  The elements of the Wilson line operator describing the transport of a Bloch state from initial quasi-momentum ${\bf k}(0)={\bf Q}$ to ${\bf k}(t)$ is measured through the overlap between the Bloch bands:
\begin{equation}
W_{Q\rightarrow {\bf k}(t)}^{mn} = \langle \Phi_{\bf k}^m | e^{i({\bf k}(t) - {\bf Q})\cdot{{\bf r}})} | \Phi_{\bf Q}^n\rangle = \langle u_{\bf k}^m | u_{\bf Q}^n \rangle.
\end{equation}
The Bloch states are defined as $|\Phi_{\bf k}^n\rangle = e^{i{\bf k}\cdot {\bf r}}|u_{\bf k}^n\rangle$. Such measurements allow for an identification of the generalized Wilczek-Zee connection ${\bf A}_{\bf k}^{n,m}=i\langle u_{\bf k}^n | {\bm\nabla}_{\bf k} | u_{\bf k}^m\rangle$, through
\begin{equation}
W_{Q\rightarrow {\bf k}(t)}^{mn} = {\cal P}\exp\left[ i\int_{\cal C} {\bf A}_{\bf k}^{m,n} d{\bf k}\right].
\end{equation}
Here, the path integral runs over the path ${\cal C}$ in reciprocal space from ${\bf Q}$ to ${\bf k}(t)$. The mixing angle $\theta({\bf k})$ is then measured from the band populations. Ramsey or Stuckelberg interferometry can be applied to measure $\phi({\bf k})$. 

\subsection{Light and Circuit Quantum Electrodynamics Architectures}

One could realize similar Floquet protocols in Circuit Quantum Electrodynamics (CQED) arrays. The Haldane model can be realized using the protocol of Eq. (19) in Ref. \cite{RevueQED} and then one could envision to build two two-dimensional cQED architectures coupled through a (small) capacitive or inductive coupling. 

Sending microwave light in the system with the appropriate frequency, one could observe counter-propagating light flows at the edges in the two arrays. 

In addition, one could test the proximity effect observed at weak-coupling through the path integral approach by coupling two qubits (two spins) on the Bloch sphere through an antiferromagnetic Ising interaction. Following the protocol of Refs. \cite{SantaBarbara,Boulder}, one could start with two radial magnetic fields ${\bf d}_1$ and ${\bf d_2}$ with a similar form as in the Letter, and fix $d^1_z = \cos\theta$ for one qubit and $d_z^2=0$ for the other. Now, by driving adiabatically the first qubit from the north to the south pole by changing the polar angle $\theta(t)=v(t-t_{in})$ from $0$ to $\pi$ at final time --- $t_{in}$ being the initial time and $v$ the speed of the protocol --- one could reconstruct the Chern number on the Bloch sphere for the first spin-1/2  \cite{SantaBarbara,Boulder,LoicPeter}. Adding an antiferromagnetic coupling between the two spins, then one could prepare the second spin in a `down' state at the north pole. When rolling the first spin to the south pole, the second spin should also flip his polarization, then transferring a Chern number for the second spin-1/2 at the end of the protocol.  Studying this protocol on the Bloch sphere as a function of the coupling between spins and speed $v$, and analyzing the behavior of the Berry curvature around the equator when including the effect of dissipation, is an interesting open question. 

\end{document}